\documentclass[aps,reprint,amsmath, amssymb,groupedaddress,floatfix]{revtex4-1}
\usepackage{natbib}
\usepackage{graphicx}
\usepackage{bm}
\usepackage{color,soul}
\usepackage{hyperref}
\usepackage{float}
\usepackage{csquotes}
\usepackage{color,soul}
\usepackage{hyperref}
\hypersetup{colorlinks=true, urlcolor=blue, citecolor=blue}

\begin{document}

\preprint{APS}
\title{Pressure and Inversion Symmetry Breaking Field Driven First Order Phase Transition and Formation of Dirac Circle in Perovskites}


\author{Ashish Kore$^1$}
\email{ These authors contributed equally to the work}
\author{Ravi Kashikar$^2$}
\email{ These authors contributed equally to the work}
 \author{Mayank Gupta$^2$}
\email{ These authors contributed equally to the work}
\author{Poorva Singh$^{1}$}
\email{ poorvasingh@phy.vnit.ac.in}
\author{B. R. K. Nanda$^2$}
\email{nandab@iitm.ac.in}
\affiliation{%
 1. Department of Physics,\\ Visvesvaraya National Institute of Technology, Nagpur-10 India \\
}%

\affiliation{
 2. Condensed Matter Theory and Computational Lab, Department of Physics,
Indian Institute of Technology Madras, Chennai - 36, India
}

\begin{abstract}
Through model Hamiltonian studies and first-principle electronic structure calculations, we have examined the effect of inversion symmetry breaking (ISB) field and hydrostatic pressure on the band topology of halide perovskites by taking MAPbI$_3$ as a prototype. Our study shows that while hydrostatic pressure induces normal to topological insulator continuous phase transition, the ISB field makes it first order. The pressure smoothly reduces the normal bandgap, and without ISB, the system achieves a gapless state before it produces a non-trivial bandgap with inverted characters. The ISB field does not stabilize the gapless state, and therefore, the discontinuity in the bandgap with pressure gives rise to the first-order transition. Furthermore, in the non-trivial phase, the ISB field forms an invariant surface Dirac circle in the neighbourhood of TRIM, which is first of its kind. The circle is formed due to interpenetration of Dirac cones resembling the band topology of AA-stacked bilayer graphene.
\end{abstract}

\maketitle

\section{ INTRODUCTION:}

In the last two decades, the phenomena of non-trivial band topology have occupied an ample space in the research area of condensed matter physics and materials science. Purely determined by the crystal and orbital symmetries through certain invariant numbers (Z$_2$), these phases exhibit unique surface/edge electronic states, other than the bulk insulating phase, invariant under adiabatic perturbations \cite{rvw2,rvw3, rvw1}. For example, in the case of topological insulators (TI), the surface/edge creates invariant conducting Dirac states.

\begin{figure}[h]
\centering
\includegraphics[angle=-0.0,origin=c,height=5.5cm,width=6.5cm]{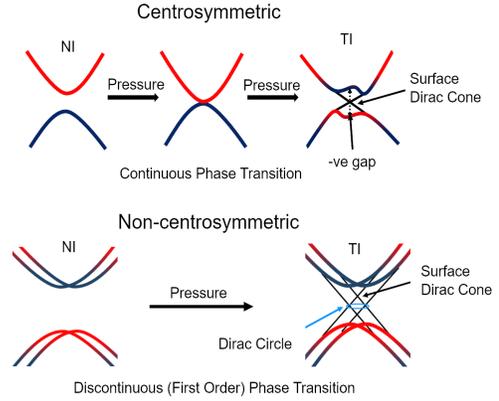}
\caption{Schematic illustration of topological phase transition in centrosymmetric and non-centrosymmetric crystal systems under hydrostatic pressure. }
\label{fig0}
\end{figure}

  Mostly alloys formed by heavy elements are being examined for exploring the non-trivial band topology in the crystalline systems. Some of the well studied examples are 3D M$_2$X$_3$ class which includes Bi$_2$Sb$_3$, Bi$_2$Se$_3$, Bi$_2$Te$_3$, Sb$_2$Te$_3$ \cite{Zhang438}, Heusler materials and half Heusler materials family \cite{Feng235121}.  However, recent studies suggest that the perovskites ABX$_3$, where A is an inorganic (e.g. Cs, Ba), B is a heavy sp-element (e.g. Bi, Sn, Pb), and X is a halogen or oxygen possess appropriate crystal and orbital symmetry to exhibit the non-trivial phases\cite{Nature_mat,BK}. Though in their equilibrium configuration, these centrosymmetric compounds are wide-bandgap insulators, recent theoretical studies predict that under hydrostatic pressure or epitaxial strain they can exhibit continuous topological phase transition (TPT)\cite{Ravi,Freeman2012}. Under continuous TPT, with compression, the bulk bandgap closes to zero at the time-reversal invariant momenta (TRIM) and then reopens with inverted characters as schematically illustrated in Fig. \ref{fig0}. As a consequence, time-reversal symmetry protected conducting states form at the surface.

The band inversion in a typical centro-symmetric TI material is driven through spin-orbit coupling (SOC) arising due to the radial electric field of the nucleus. However, if the inversion symmetry is broken a new dipolar electric field emerges giving rise to Rashba coupling\cite{rashba_soc_1, DFT_non_centro}. This inversion symmetry breaking (ISB) field has profound effect in creating complex conduction and valence band structures leading to their applications in optoelectronics and thermoelectrics \cite{non_centr1,non_centr2}. Among the ABX$_3$ perovskite oxides and halides, the halides can exist in both centrosymmteric and non-centrosymmetric depending upon the entity A and crystal phase. First principle calculations show that if A is an organic molecule such as CH$_3$NH$_3$ (MA), the inversion symmetry is broken \cite{DFT_non_centro}. At the same time, the experimental studies depending on the growth conditions and characterization techniques predict the organic halides such as MAPbI$_3$ to be either centro-symmetric \cite{non_centre_d2} or non-centrosymmetric \cite{non_centro_2, Beecher, non_centro}. Therefore, we find MAPbI$_3$ as a prototype compound where the effect of ISB field on the electronic structure can be theoretically proposed and experimentally verified.  The other reason behind choosing MAPbI$_3$ as a prototype is that experimentally, high pressure (up to 60 GPa) studies are carried out on this compound and metallization is demonstrated which is attributed to either the partial amorphoziation of the compound or due to topological insulator driven metallic surface state or both\cite{pressure_metal}.

 The objective of this work is to study the evolution of the band topology of MAPbI$_3$ with pressure and ISB field and for this purpose we have employed  density-functional calculations and Slater-Koster formalism based tight-binding (SK-TB) Hamiltonian studies. We find that unlike the continuous TPT, the ISB in MAPbI$_3$ brings a first order quantum phase transition with pressure as schematically illustrated in the lower panel of Fig. \ref{fig0}. Here, the bulk bandgap gradually decreases, but before it closes to zero, a negative bandgap with inverted character appears. 
Such a discontinuous TPT is fundamentally significant and also important in designing topological devices where the topology can be tuned through pressure or strain.  So far the discontinuous TPT has only been observed through chemical doping in two intermetallic alloys (e.g. Pb$_{1-x}$Sn$_x$Se and TlBi(S$_{1-x}$Se$_x$) \cite{TPT1,TPT2,TPT3,TPT4}. However, realizing them in MAPbI$_3$ with pressure brings an additional multifunctional dimensionality to these promising photovoltaic family of members. Furthermore, concerning the surface band structure, the ISB field breaks the four-fold degenerate point node on the Fermi level at the TRIM and instead create two twofold degenerate point nodes, one higher in energy and other lower in energy with a Dirac circle in between. The radius of the circle increases with the ISB field.Thus the surface band topology of the non-centrosymmetric perovskite under pressure can reproduce the band topology of AA-stacked bilayer graphene \cite{Dirac_cone}. The ISB field strength plays the role of interlayer-coupling in the graphene bilayer.

\section{Structure and Computational Details:}

\begin{figure}[h]
\centering
\includegraphics[angle=-0.0,origin=c,height=8cm,width=7.5cm]{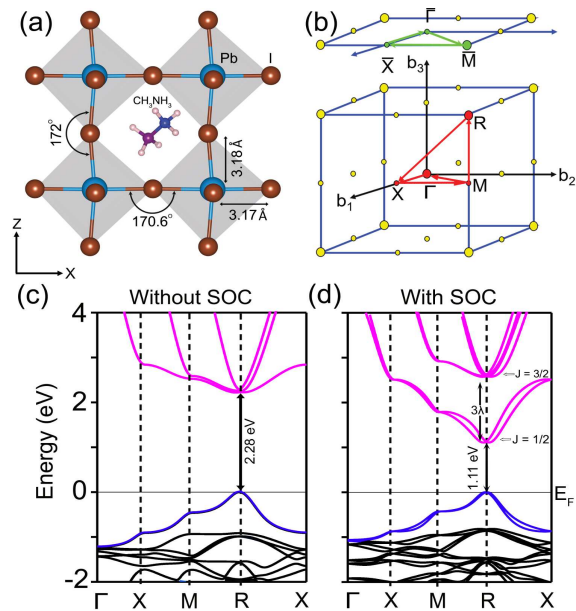}
\caption{(a) Crystal structure of  non-centrosymmetric MAPbI$_3$. The organic cation CH$_3$NH$_3^+$ breaks the inversion symmetry. (b) Corresponding bulk and surface Brillouin zone with high symmetry $k$-points.  (c) and (d) The bulk band structure of MAPbI$_3$ in absence and presence of SOC. The SOC lifts the degeneracy in the energy domain due to atomic effect and in the momentum space due to Rashba effect driven by non-centrosymmetricity.}
\label{fig1}
\end{figure}

Close to room temperature  ($\sim$ 327K) MAPbI$_3$ stabilizes in the cubic phase (space group Pm-3m), but with a T$_{2u}$ mode of distortion to the octahedra as shown in Fig. \ref{fig1}(a) \cite{struct}.  However, exactly at the room temperature, the lattice distorts through minor rotation to the octahedra which results in a pseudocubic structure. The structural relaxation of MAPbI$_3$ through DFT methods breaks the inversion symmetry to make the crystal non-centrosymmetric \cite{DFT_non_centro}. It has been shown that such minor rotations do not adversely affect the band topology of the crystal and the cubic configuration, as considered in this paper, is sufficient to study the band phenomena of this compound. As the objective is to examine the  electronic structure of MAPbI$_3$ as a function of pressure, we carried out band structure calculations for a discrete set of (V/V$_0$), where V$_0$ is the equilibrium volume corresponding to DFT optimized lattice parameter of 6.44 \AA which slightly overestimates the experimental value of 6.33 \AA. 

The band structures are calculated using Full potential based linearized augmented plane wave (FP-LAPW )methods as implemented in WIEN2k code  \cite{Blaha}. For the description of exchange correlation potentials, the PBE formalism for generalized gradient approximation combined with modified Becke-Johnson (mBJ) correction is used \cite{GGA,mbj-1,mbj-2}. The
largest vector in the plane wave expansion is obtained by
setting R$_{MT}$K$_{max}$ to 3.0. A 6$\times$6$\times$6 $k$-mesh, yields 112 irreducible $k$-points, is used for the Brillouin zone integration. Self-consistent calculations are carried out with augmented plane waves of interstitial regions and localized orbitals (6s and 6p of Pb, 5p of I) within the muffin-tin spheres (R$_{MT}^{C}$ = 1.33 a.u., R$_{MT}^{H}$ = 0.68 a.u., R$_{MT}^{N}$ = 1.26 a.u., and R$_{MT}^{Pb}$ = R$_{MT}^{I}$ = 2.5 a.u.). A 25 unit cell thick slab grown along 001 is used to calculate the surface band structure using both Wannier functions based TB formalism as implemented in wanniertools \cite{wannier90,arpes} and Slater Koster based TB methods \cite{slater}. The topological invariants Z$_2$, which provides a quantitative measure of the band topology, are estimated through Wannier function based  Wilson loop method \cite{arpes}.

\section{ RESULTS AND DISCUSSION:}

\subsection{ Electronic Structure of  CH$_3$NH$_3$PbI$_3$ }

The band structure of MAPbI$_3$ arising due to chemical bonding, manifested through interactions among the valence electrons, and  due to both chemical bonding and SOC are shown in Fig. \ref{fig1}c and d respectively. When the chemical bonding alone is considered, the system shows a direct bandgap with the conduction band minimum (CBM), formed by the Pb-p orbitals (magenta) and valence band maximum (VBM), dominated by Pb-s orbital (blue) lying at the TRIM R. Such a band structure belongs to the universal class of ABX$_3$ family \cite{Ravi,ravi2}. Here, the B-\{s, p\} -X-p covalent interaction gives rise to a set of bonding bands far below the Fermi level (E$_F$) and a set of antibonding bands in the vicinity of E$_F$. In the antibonding spectrum there is a bandgap between the lower lying B-s dominated band and the upper lying B-p dominated bands. If the valence electron count (VEC) is less than 18 (e.g. KBiO$_3$), all the antibonding bands are unoccupied. If VEC is 20, the B-s dominated antibonding band is occupied while the B-p dominated antibonding bands are empty to form the bandgap at E$_F$ (e.g. CsSnI$_3$ and CsPbI$_3$)\cite{Ravi,BK}. Same is the case here as the MAPbI$_3$ contributes 20 valence electrons to the system. Beyond the nearest neighbor Pb-I coupling, we shall see below through a tight-binding model that the band topology is very sensitive to second neighbor Pb-Pb coupling. 

The SOC, arising from Pb,  makes the  band structure complex at the Fermi surface. The SOC is introduced to the Hamiltonian of the system through $\hat{H} = -\frac{\mu_B}{2mc}\Vec{\sigma}\cdot(\Vec{E}(r) \times \Vec{p} )$. Where $\sigma$ is the spin of the electron occupying a given state and $\Vec{E}$ is the field experienced by the electron. In MAPbI$_3$ there are two kinds of electric field that manipulate the SOC. The first one is the radial electric field emerging from the Pb$^{2+}$ cation and the second one is due to the presence of a polarizing field in this non-centrosymmetric system. While, the former is widely known as the \textbf{L.S} coupling, the latter is considered to be Rashba coupling. The \textbf{L.S} coupling lifts the energy degeneracy between the J=1/2 and 3/2 states as can be seen from Fig. \ref{fig1}d. The Rashba coupling lifts the momentum degeneracy in the whole Brillouin zone except the TRIM.  The TB model described below provides the intricate details of the role of Pb-Pb second neighbor interactions, \textbf{L.S} coupling and Rashba coupling on the band topology of MAPbI$_3$.

\subsection{Tight-Binding Model}
The TB model Hamiltonian that can appropriately explain the band structure of the non-centrosymmetric halide perovskite is as follows.

\begin{align}
     H& = H_{TB} + H_{atomic} + H_{ISB} \nonumber\\ 
      & =   \sum_{i,\alpha}\epsilon_{i\alpha}c_{i\alpha}^\dag c_{i\alpha} + \sum_{ij;\alpha,\beta}t_{i\alpha j\beta}(c_{i\alpha}^\dag c_{j\beta} + h.c)+ \nonumber \\
      &\lambda\textbf{L}\cdot\textbf{S} + H_{ISB}
\end{align}

The first two terms in Eq. 1 are the non-SOC terms and they contribute to the formation of the band through covalent interactions among the valence states. While the first term is the on-site term with $\epsilon_{i\alpha}$ as onsite energy of the $\alpha$ orbital (Pb-\{s,p\}) at the $i$-th site, the second term represents the  second-neighbor electron hopping among the Pb-\{s, p\} states with $t$ being the hopping interaction strength. The third term is the atomistic SOC with coupling strength $\lambda$ and the fourth term is the inversion symmetry breaking field term arising from the polarized electric field  induced through the breakdown of the inversion symmetry\cite{rashba_soc_1}. This term is also known as Rashba SOC.  This field allows intermixing of the orbitals. For example the field along $\hat{x}$ allows the intermixing of the bands \{s, p$_x$\}, \{p$_y$, p$_x$\} and \{p$_z$, p$_x$\}. The general form of ISB term can be written as 

\begin{eqnarray}
H_{ISB} = \sum_{\alpha \beta, ij} \gamma_{\alpha \beta}  e^{ik\cdot(R_i - R_j)}\\
    \gamma_{\alpha \beta}^{x/y/z} = \langle \alpha, R_i |E_{x/y/z} (x/y/z)| \beta, R_j\rangle
\end{eqnarray}
If we ignore the coupling of Pb-\{s, p\} with the I-p dominated bands and consider only the Pb-Pb nearest neighbor mixing, the ISB component of the Hamiltonian in the matrix form with the basis set in the order $|s\rangle$, $|p_x\rangle$, $|p_y\rangle$ and $|p_z\rangle$ can be given by  
\begin{widetext}

\begin{equation}
    H_R = \left(\begin{smallmatrix} 0 & \gamma_{sp}^x (1-l^2) - lm \gamma_{sp}^y - ln\gamma_{sp}^z & \gamma_{sp}^y (1-m^2) - lm \gamma_{sp}^x - mn\gamma_{sp}^z  & \gamma_{sp}^z (1-n^2) - ln \gamma_{sp}^x - mn\gamma_{sp}^y  \\
    \gamma_{sp}^x (1-l^2) - lm \gamma_{sp}^y - ln\gamma_{sp}^z 
    &0 
    &\gamma_{pp}^yl-\gamma_{pp}^xm   
    & \gamma_{pp}^zl - \gamma_{pp}^xn \\
    \gamma_{sp}^y (1-m^2) - lm \gamma_{sp}^x - mn\gamma_{sp}^z 
    & \gamma_{pp}^xm - \gamma_{pp}^yl 
    &0
    &\gamma_{pp}^zm - \gamma_{pp}^yn  \\
    \gamma_{sp}^z (1-n^2) -  ln\gamma_{sp}^x - mn\gamma_{sp}^y 
    & \gamma_{pp}^xn - \gamma_{pp}^zl
    & \gamma_{pp}^yn - \gamma_{pp}^zm  
    &0 \\ \end{smallmatrix}\right)
\end{equation}
Here, $l$, $m$, and $n$ are the direction cosines connecting $R_j$ with $R_i$. 
    
\begin{equation}
    \mathbf{H_R} = 
    \left(\begin{smallmatrix}
 0& 2\gamma_{sp}^x(C_y+C_z)& 2\gamma_{sp}^y(C_z+C_x)& 2\gamma_{sp}^z(C_x+C_y)\\
    2\gamma_{sp}^x(C_y+C_z)& 0& 2i(\gamma_{pp}^y S_x-\gamma_{pp}^x S_y)& -2i(\gamma_{pp}^xS_z-\gamma_{pp}^zS_x)\\
    2\gamma_{sp}^y(C_z+C_x)& -2i(\gamma_{pp}^y S_x- \gamma_{pp}^x S_y)& 0& -2i(\gamma_{pp}^y S_z-\gamma_{pp}^z S_y)\\ 
    2\gamma_{sp}^z(C_x+C_y)& 2i(\gamma_{pp}^xS_z-\gamma_{pp}^zS_x)& 2i(\gamma_{pp}^y S_z- \gamma_{pp}^z S_y)& 0\\
       \end{smallmatrix}\right);  S_x = sin(k_xa), C_x = cos(k_xa)
\end{equation}
\end{widetext}
Materials having both time reversal as well as inversion symmetries lead to  double degenerate energy eigenvalues at each $k$-point, which can be formulated as $E(k,\uparrow) = E(-k,\downarrow)$ and $
E(k,\uparrow) = E(-k,\uparrow) $. Therefore, the broken inversion symmetry lifts the doubly degenerate bands of electronic spectrum other than TRIM due to Rashba SOC. In addition to this, the TB and atomic SOC Hamiltonian matrices are given by,  
\[ 
\mathbf{H_{O}+H_{atomic}}
 =  \left( \begin{array}{cc}
    \mathbf{H_{\uparrow\uparrow}}& \mathbf{H_{\uparrow\downarrow}} \\
    \mathbf{H_{\downarrow\uparrow}^{\dagger}}&  \mathbf{H_{\downarrow\downarrow}}
\end{array}  \right), 
H_{\uparrow\downarrow}
 =  \left( \begin{array}{cccc}
    0&0&0&0\\
    0&0&0&\lambda\\ 
    0&0&0&-i\lambda\\
    0&\lambda&-i\lambda&0
\end{array}  \right),
\]
\[
\mathbf{ H_{\uparrow\uparrow}}= \mathbf{(H_{\downarrow\downarrow})^{\dagger}}
 =  \left( \begin{array}{cccc}
    \epsilon_s+f_0 & 2it_{sp}S_x & 2it_{sp}S_y & 2it_{sp}S_z\\
    -2it_{sp}S_x & \epsilon_p+f_1 & -i\lambda &0\\ 
    -2it_{sp}S_y & i\lambda & \epsilon_p+f_2 & 0\\
    -2it_{sp}S_z &   0 & 0 & \epsilon_p+f_3 
\end{array}  \right)
\]

Here, $\epsilon_s$ and $\epsilon_p$ are effective on-site energies and $'t'$s are hopping interactions between different orbitals and, 
\begin{eqnarray}
f_0& = &2t_{ss}(C_x+C_y+C_z) \nonumber\\
f_1& = &2t_{pp\sigma}C_x+2t_{pp\pi}(C_y+C_z) \nonumber \\
f_2& = &2t_{pp\sigma}C_y+2t_{pp\pi}(C_x+C_z) \nonumber\\
f_3& = &2t_{pp\sigma}C_z+2t_{pp\pi}(C_x+C_y).
\end{eqnarray}
The eigenvalues are obtained from the exact diagonalization along the high symmetry $k$-path and are fitted with DFT bands to estimate the on-site energies, hopping parameters, and SOC and they are listed in Table-I. The variations of these TB parameters with uniform compression, which we have used in this work as an external stimuli to study the band phenomena in pervoskites, are provided in supplementary material.

\begin{table*}
\centering
\caption{ On-site and hopping parameters of MAPbI$_3$ in unit of eV. }
\begin{tabular}{ccccccccccccc}
\hline
$\epsilon_{s}$ &$\epsilon_{p}$ & t$_{ss}$ & $t_{sp\sigma}$&$t_{pp\sigma}$&$t_{pp\pi}$ & $\lambda$  & $\gamma_{sp}^x$ & $\gamma_{sp}^y$ & $\gamma_{sp}^z$ & $\gamma_{pp}^x$ & $\gamma_{pp}^y$ & $\gamma_{pp}^z$\\
 \hline
 -0.62	&3.96&	-0.11 &	0.4 &	0.77 &	0.08& 0.49 & 0.045 &0.0&0.03&-0.045&0.12&0.06 \\
 \hline
\end{tabular}
\label{T3}
\end{table*}

\subsection{Bandgap discontinuity and signature of first order phase transition}

\begin{figure}
\centering
\includegraphics[angle=-0.0,origin=c,height=15cm,width=8cm]{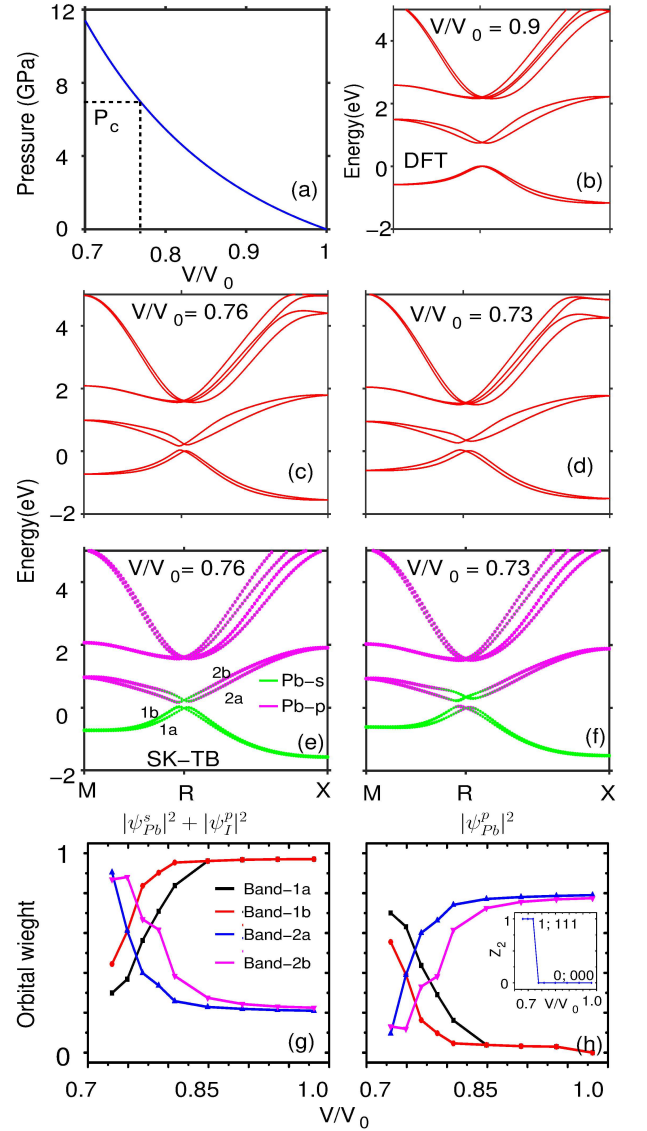}
\caption{(a) Pressure vs. volume curve for MAPbI$_3$, as obtained from Birch-Murnaghan equation of state. Here, the P$_c$ corresponds to critical pressure at which normal insulator (NI) to topological insulator (TI) transition occurs. (b-d) DFT obtained band structure of MAPbI$_3$ for different V/V$_0$. (e-f) Orbital resolved  TB band structure of MAPbI$_3$ at and just below the critical compression. The non-trivial phases is characterized by the band inversion at the Fermi level. Thus, the bandgap in non-trivial phases is known to be as negative bandgap. (g-h) Variation of  Pb-\{s, p\} and I-p orbital character of band-1 and 2 as a function of V/V$_0$. The inset in (h) shows the variation of topological invariant quantity Z$_2$ as a function of V/V$_0$.}
\label{fig2}
\end{figure}

\begin{figure}
\centering
\includegraphics[angle=-0.0,origin=c,height=4.5cm,width=8.5cm]{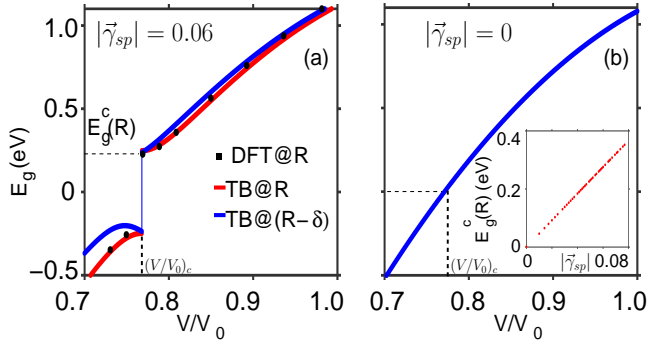}
\caption{ (a) The discontinuous bandgap variation of MAPbI$_3$ in TPT  under hydrostatic pressure as obtained from DFT data and simulated with TB model for the resulted ISB field ($|\gamma_{sp}^z|$). (b) The variation in the bandgap become continuous, when the ISB field is set to zero. The inset shows the variation of critical bandgap as function of ISB field. }
\label{fig3}
\end{figure}

In this section we investigate the evolution of the band topology with compression which can be achieved experimentally by applying hydrostatic pressure. Fig. \ref{fig2}a plots the required applied pressure as a function of compression which is obtained by employing the second order Birch–Murnaghan equation of state with the DFT estimated bulk modulus of 15.64 GPa (the experimental value is 12-16 GPa )\cite{BM1, BM2, bulk_m}. Fig. \ref{fig2}(b-d) demonstrate the change in the band structure with compression for three representative values of V/V$_0$.  These three figure infer that with compression, the bandgap initially decreases (see the case of V/V$_0$ = 0.9), reaches a minimum (around V/V$_0$ = 0.76) and on further compression there is a reopening of the gap (see the case of V/V$_0$ = 0.73) which provide the first indication of a TPT under compression that can be achieved by applying a pressure of 7 GPa as Fig. \ref{fig2}a suggests. It may be noted that there are experimental report showing reasonable crystallinity can be maintained in MAPbI$_3$ with applied pressure up to 60 GPa \cite{pressure_metal}. The same report also hints towards a TPT. 

To further confirm the normal to topological phase transition, the first step is to examine the s-p band inversion at TRIM $R$ of the Brillouin zone. The Pb-$\{$s,p$\}$ orbital projected SK-TB band structure, which is in excellent agreement with the all electron DFT band structure, are shown in Fig. \ref{fig2}(e-f). For V/V$_0$ = 0.76, the inversion of the $s-$ and $p-$ characters between the lower bands (1a, 1b) and the upper bands (2a, 2b) at the $R$ is about to happen. At V/V$_0$ = 0.73, a complete s-p bands inversion is observed. To provide a quantitative estimation of the band inversion, in Fig. \ref{fig2}(g-h) we have plotted the orbital projected charge densities ($|\Psi_s^{Pb}(k)|^2+|\Psi_p^{I}(k)|^2$ and $|\Psi_p^{Pb}(k)|^2$ ) at R as function of compression. The selection of the orbital is based on the fact that the bands (1a, 1b) are dominated by both Pb-s and I-p characters and bands (2a, 2b) are dominated by the Pb-p characters as a result of Pb-$\{$s, p$\}$-I-p hybridization. From the figure we gather that below the critical compression (V/V$_0$ $>$ 0.76), at R the band 1a and 1b are formed by Pb-s and I-p states while band 2a and 2b are formed by Pb-p states. However, the inverse happens above the critical compression. The value of topological invariants Z$_2$ as a function of compression (see inset Fig. \ref{fig2}h) concurs the NI-TI phase transition with band inversion. 

Further understanding of the evolution of the band topology is garnered by examining the compression dependent bandgap at R and around its neighborhood R$\pm \delta$ which are plotted in Fig. \ref{fig3}. The DFT obtained gap ( Fig. \ref{fig3}a) smoothly decreases and reaches a finite minimum at the critical compression. On further compression a new negative bandgap emerge with VBM and CBM inverting their character (see Fig. \ref{fig2}f and the caption for definition of the negative bandgap). This is in contrast to a centro-symmetric system such as CsSnI$_3$ where on compression the bandgap gradually vanishes to zero to create an accidental Dirac semimetal phase before reopening the negative bandgap\cite{Ravi}. 

To understand the cause of discontinuity in the variation of the band gap with compression, we employed the TB model and examined the nature of bandgap as a function of ISB field through the Rashba coupling strength $\gamma$ (see Eq. 3). We find that while $\gamma_{pp}$ mainly contribute towards splitting of the bands (1 - 4) in the momentum space, they hardly affect the bands in the energy domain. On the other hand, $\gamma_{sp}$ affects the bands both in the momentum and energy domain. Its effect on the bandgap can be understood from Fig. \ref{fig3}. For the optimized value of $\gamma_{sp}$ ( = 0.06 eV), the TB obtained E$_g$ matches well with that of the DFT (see Fig. \ref{fig3}a) and reproduces the bandgap discontinuity.  However, if $\gamma_{sp}$ is set to zero, as in the case of centro-symmetric system, the discontinuity vanishes. Fig. \ref{fig3}b inset, where the minimum value of E$_g$ at $R$ and at the critical compression is plotted as a function of $\gamma_{sp}$, shows that any finite value of $\gamma_{sp}$ introduces the discontinuity which has not been observed so far in the family of perovskites. As of now, such phase transitions, analogous to first order, where the gapless electronic state cannot be stabilized, are experimentally reported in two of the Se based alloys, namely, (Pb, Sn)Se \cite{TPT1,TPT2,TPT3,TPT4}.

\subsection{Surface electronic structure} 

 \begin{figure}
\centering
\includegraphics[angle=-0.0,origin=c,height=8.3cm,width=9.0cm]{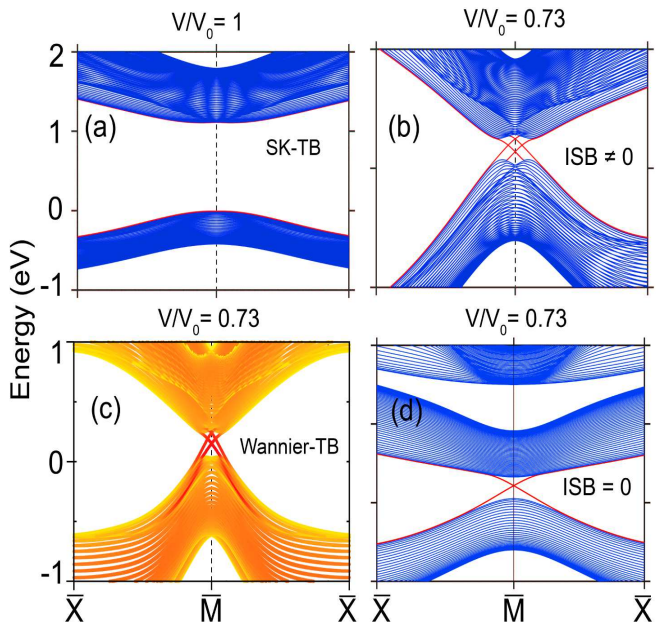}
\caption{Surface TB band structure of MAPbI$_3$ for (a) equilibrium structure and (b) below critical compression. (c) Same as (b) but obtained through Wannier based TB formalism. (d) Surface band structure in the absence of ISB field.}
\label{fig4}
\end{figure}

 The band inversion and estimation of Z$_2$ as 1 below a critical compression implies that there will be symmetry protected robust surface states in this compound. To examine and verify the formation of these states, we have applied the TB model on a slab of 25 unit cell thick grown along 001 direction and terminated with PbI$_2$ layer. The slab Hamiltonian, where the compression dependent various coupling constants are used as input, is described in the supplementary information. The resulted surface states are shown in Fig. \ref{fig4}. In the absence of compression (V/V$_0$ = 1), where the bulk compound is a normal insulator, the surface electronic structure also produces a gap as expected. However, as the compound is compressed below the critical value (e.g. V/V$_0$ = 0.73), the electronic structure yields conducting states around $\bar{M}$. Along the path $\bar{X}-\bar{M}-\bar{X}$, two such conducting states are observed at $\bar{M}\pm \delta$. Which is further reconfirmed using the standard Wannier calculations as shown in Fig. \ref{fig4}(c). However, if the ISB field is set to zero as in the case of centro-symmetric compound, the linearly dispersed valence and conduction bands touch each other only at $\bar{M}$ replicating the conventional topological insulators.
 
\begin{figure}[h]
\centering
\includegraphics[angle=-0.0,origin=c,height=9.0cm,width=7.5cm]{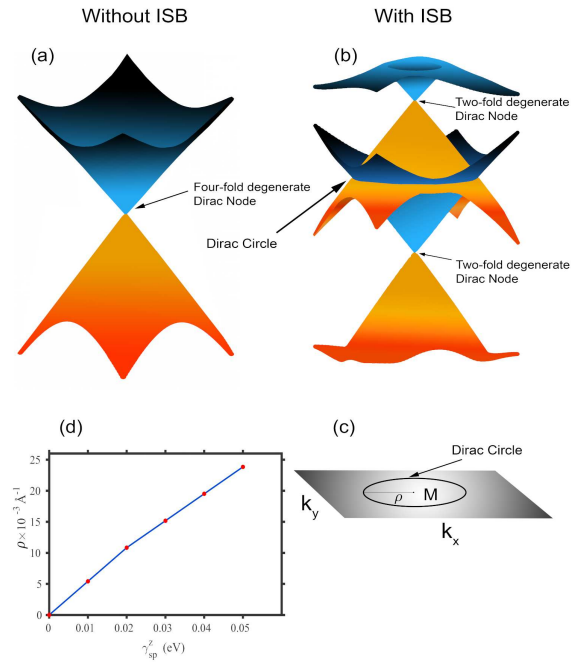}
\caption{The 3D band structure, (a) without and (b) with  the ISB field, plotted on the k$_x$ - k$_y$ plane around the high symmetry point $\bar{M}$ of the surface Brillouin zone. (c) Two dimensional projection of Dirac circle in $k_x$-$k_y$ plane. (d) Radius($\rho$) of the Dirac circle as a function of the coupling strength ($\gamma_{sp}^z$).}
\label{fig5}
\end{figure}

 To further elucidate the non-trivial surface electronic structure we have moved from a single path in the surface Brillouin zone to the plane in the neighborhood of $\bar{M}$ and the resulted band structures are demonstrated in Fig. \ref{fig5}. In the absence of ISB field, it produces a two doubly degenerate  Dirac cone, one each from valence and conduction bands. The degeneracy appears due to the Kramer pair formation. The Dirac cones touch each other at $\bar{M}$ and are protected by time reversal symmetry. Therefore, the centrosymmetric system gives rise to a fourfold-degenerate point node at $\bar{M}$. With the ISB field, the Kramer pairing breaks and as a consequence, one pair of Dirac cones -- one each from valence band and conduction band --  shifts upwards and the other pair shifts downwards in energy as shown in Fig. \ref{fig5}b. Thereby, the fourfold degenerate point node breakdown to two twofold degenerate point nodes, one higher in energy and other lower in energy with a Dirac circle lying in between as shown in Fig. \ref{fig5}d. 
 The formation of two-fold degenerate point nodes and a Dirac circle are yet to be found in the broad family of topological insulators. However, this surface band topology completely imitate the band structure of AA-stacked graphene\cite{Dirac_cone} and $\gamma_{sp}^z$ plays the role of inter-layer coupling strength.
 
 In conclusion, we demonstrate that the inversion symmetry breaking field arising from the presence of the organic molecule CH$_3$NH$_3$ does not allow the halide perovskite to stabilize a gapless state  and thereby induces first order normal to topological phase transition with pressure in this compound. Such phase transition is rare and was earlier seen in two of the intermetallic alloys (Pb$_{1-x}$Sn$_x$Se, TlBiS$_{1-x}$Se$_x$).  Furthermore, non-trivial surface states that emerge below a critical compression of the compound consist of two-fold degenerate Dirac point nodes -- one below the Fermi level and the other above it-- and one Dirac circle lying exactly on the surface Fermi level. While it replicates the band structure of AA stacked bilayer graphene around the Dirac point K, it is of its first kind among the family of 3D topological insulators.\\\\   
 \textbf{ACKNOWLEDGMENT:} Author PS would like to thank NPSF C-DAC Pune for providing HPC facility. BRKN acknowledges the computational resources provided by HPCE, IIT Madras and Department of Science and Technology, India  for funding through Grant No. EMR/2016/003791.\\\\

\bibliography{paper}%

\providecommand{\noopsort}[1]{}\providecommand{\singleletter}[1]{#1}%

\end{document}